\documentclass{article}
\usepackage[preprint]{spconf}
\copyrightnotice{\copyright\ IEEE 2020}
\usepackage{color,xcolor}
\usepackage{epsfig}
\usepackage{graphicx}

\usepackage{array}
\usepackage{booktabs}
\usepackage{colortbl}
\usepackage{multirow}
\usepackage{float}
\usepackage{caption}
\usepackage[labelformat=simple]{subcaption}

\usepackage{footnote}
\makesavenoteenv{tabular}
\makesavenoteenv{table}

\usepackage{amsmath,amsfonts,amssymb,bm}
\usepackage[super]{nth}

\usepackage{changepage}
\usepackage{extramarks}
\usepackage{fancyhdr}
\usepackage{lastpage}
\usepackage{setspace}
\usepackage{soul}
\usepackage{xspace}

\usepackage{url}
\usepackage{hyperref}
\hypersetup{colorlinks=True, urlcolor=black}
\usepackage[numbers,sort&compress]{natbib}
\usepackage{algorithm, algorithmic}
\usepackage{enumitem}
\usepackage{verbatim}
\usepackage{pifont}
\usepackage[acronyms]{glossaries}
\glsdisablehyper

\newcommand{\sect}[1]{Section~\ref{sec:#1}}
\newcommand{\sectdot}[1]{Sec.~\ref{sec:#1}}

\newcommand{\ssectdot}[1]{Sec.~\ref{ssec:#1}}

\newcommand{\figdot}[1]{Fig.~\ref{fig:#1}}
\newcommand{\tbl}[1]{Table~\ref{tab:#1}}

\newcommand{\twosect}[2]{Sections~\ref{sec:#1} and \ref{sec:#2}}

\newcommand{\ignore}[1]{}

\makeatletter
\DeclareRobustCommand\onedot{\futurelet\@let@token\@onedot}
\def\@onedot{\ifx\@let@token.\else.\null\fi\xspace}

\def\eg{\emph{e.g}\onedot} 
\def\ie{\emph{i.e}\onedot}

\def\wrt{w.r.t\onedot}

\makeatother

\definecolor{MyDarkBlue}{rgb}{0,0.08,1}
\definecolor{MyDarkGreen}{rgb}{0.02,0.6,0.02}
\definecolor{MyDarkRed}{rgb}{0.8,0.02,0.02}
\definecolor{MyDarkOrange}{rgb}{0.40,0.2,0.02}
\definecolor{MyPurple}{RGB}{111,0,255}
\definecolor{MyRed}{rgb}{1.0,0.0,0.0}
\definecolor{MyGold}{rgb}{0.75,0.6,0.12}
\definecolor{MyDarkgray}{rgb}{0.66, 0.66, 0.66}

\def\presec{\vspace{-0.3em}}
\def\postsec{\vspace{-0.3em}}

\title{Confidence Estimation for Attention-based\\Sequence-to-sequence Models for Speech Recognition}
\name{\begin{tabular}{c}
      Qiujia Li$^{1*}$, David Qiu$^2$, Yu Zhang$^2$, Bo Li$^2$, Yanzhang He$^2$,\\
     Philip C. Woodland$^1$, Liangliang Cao$^2$, Trevor Strohman$^2$\thanks{$^*$Work was done while the author interned at Google.}
\end{tabular}}
\address{$^1$ University of Cambridge, UK, $^2$ Google LLC, USA\\
\footnotesize{$^1$\texttt{\{ql264,pcw\}@eng.cam.ac.uk}, $^2$\texttt{\{qdavid,ngyuzh,boboli,yanzhanghe,llcao,strohman\}@google.com}}\vspace{-1em}}

\begin{document}
\ninept
\maketitle

\begin{abstract}
For various speech-related tasks, confidence scores from a speech recogniser are a useful measure to assess the quality of transcriptions. In traditional hidden Markov model-based automatic speech recognition (ASR) systems, confidence scores can be reliably obtained from word posteriors in decoding lattices. However, for an ASR system with an auto-regressive decoder, such as an attention-based sequence-to-sequence model, computing word posteriors is difficult. An obvious alternative is to use the decoder softmax probability as the model confidence. 
In this paper, we first examine how some commonly used regularisation methods influence the softmax-based confidence scores and study the overconfident behaviour of end-to-end models. Then we propose a lightweight and effective approach named confidence estimation module (CEM) on top of an existing end-to-end ASR model. Experiments on LibriSpeech show that CEM can mitigate the overconfidence problem and can produce more reliable confidence scores with and without shallow fusion of a language model. Further analysis shows that CEM generalises well to speech from a moderately mismatched domain and can potentially improve downstream tasks such as semi-supervised learning.


\end{abstract}

\presec
\section{Introduction}
\postsec
Confidence scores have been an intrinsic part of automatic speech recognition (ASR) systems~\cite{Wessel2001ConfidenceMF,Jiang2005ConfidenceMF,Yu2011CalibrationOC}. Many speech-related applications depend on high-quality confidence scores to mitigate error from speech recognisers. For example, in semi-supervised learning and active learning, utterances with highly confident hypotheses are selected to further improve ASR performance~\cite{Chan2004ImprovingBN,Tr2005CombiningAA,Riccardi2005ActiveLT}. Confidence scores are also used in dialogue systems where queries with low confidence may be returned to users for clarification~\cite{Tr2005CombiningAA}. As an indication of ASR uncertainty, confidence scores can play an role in speaker adaptation\cite{Uebel2001SpeakerAU}, and system combination~\cite{Evermann2000PosteriorPD}.

In conventional HMM-based systems, reliable confidence scores can be easily obtained by computing word posterior probabilities from compact representations of the hypotheses space, \eg lattices or confusion networks~\cite{Mangu2000FindingCI,Evermann2000PosteriorPD}. Improved confidence estimation can be achieved by using model-based approaches, such as conditional random fields\cite{Seigel2011CombiningIS}, recurrent neural networks~\cite{Kalgaonkar2015EstimatingCS,Ragni2018ConfidenceEA} and graph neural networks~\cite{Li2019BidirectionalLR}, or leveraging more related information including phonetics, word/phone duration and language models~\cite{Jiang2005ConfidenceMF,Kastanos2020ConfidenceEF}.

Recently, end-to-end speech recognition has achieved promising performance over the conventional systems~\cite{Chiu2018StateoftheArtSR}. ``End-to-end systems'' refers to end-to-end differentiable and trainable neural networks, in contrast to modular systems with separate acoustic models, language models and token passing decoders. As end-to-end speech recognition has various modelling and engineering advantages, they are becoming more widely adopted~\cite{He2019StreamingES}. One class of the end-to-end systems is attention-based sequence-to-sequence models~\cite{Chorowski2015AttentionBasedMF,Chan2016ListenAA}. The auto-regressive decoder in end-to-end systems implies that compact representations like lattices cannot be constructed for attention-based models, except using specific decoder architectures and heuristics~\cite{Zapotoczny2019LatticeGI}. Consequently, computing word posteriors in the end-to-end hypothesis space becomes prohibitively expensive. A greedy approximation would be taking the softmax probability from each step of the decoder as the confidence scores for each token~\cite{Park2020ImprovedNS}. However, the quality of confidence estimation by softmax probabilities may be very poor~\cite{Hendrycks2017ABF}. 
\vspace{-1em}
\begin{figure}[ht]
    \centering
    \includegraphics[width=0.65\linewidth]{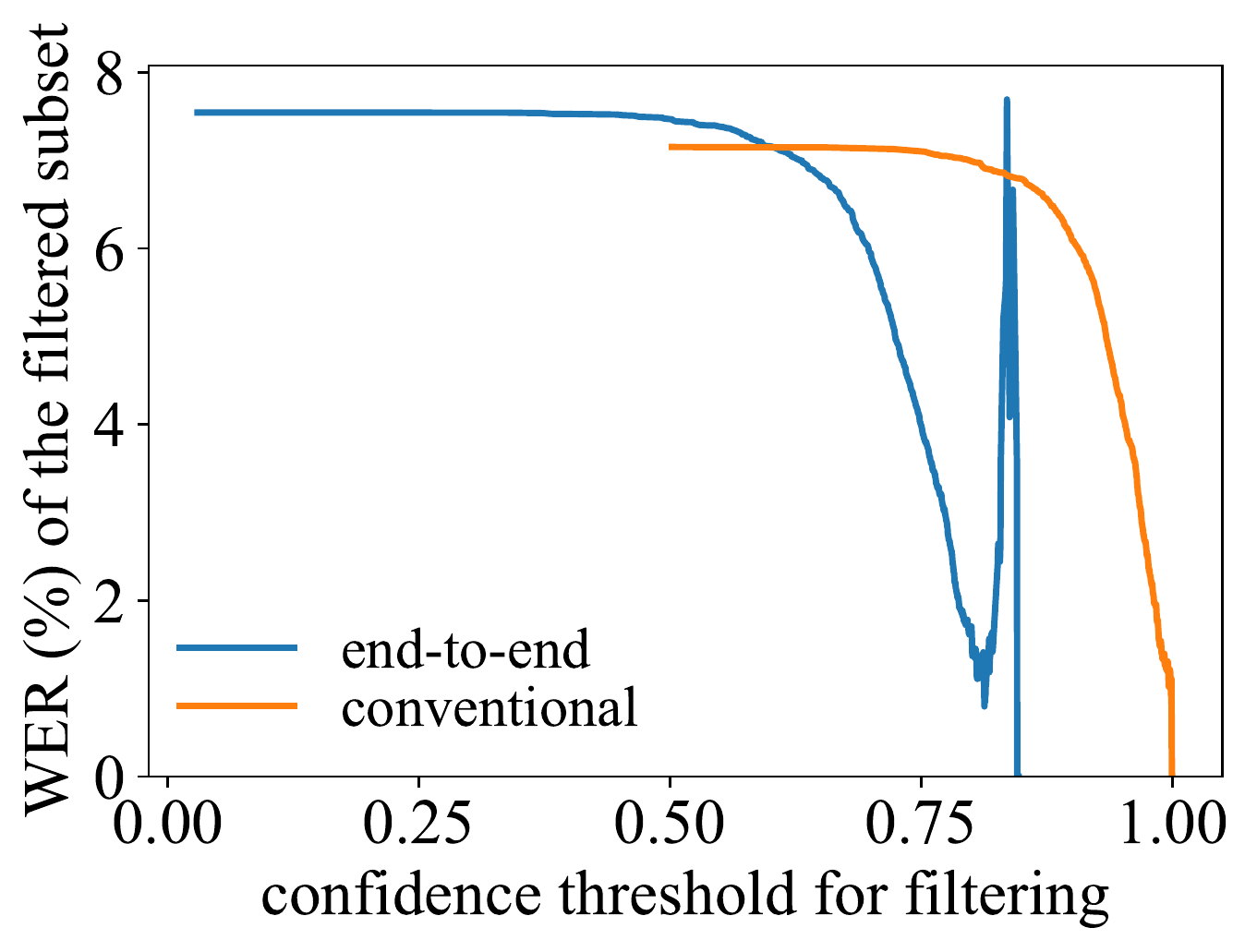}
    \vspace{-1em}
    \caption{Filtering behaviour of a conventional HMM-based system and an attention-based sequence-to-sequence model based on confidence scores. Utterances with confidence higher than a threshold (x-axis) are selected, and word error rate (WER) of the filtered subset (y-axis) are plotted. Both systems are trained on LibriSpeech 100-hour data and the test-clean set is used for filtering.}
    \label{fig:teaser}
    \vspace{-1em}
\end{figure}

This issue is illustrated by \figdot{teaser} which is a plot used for data selection for semi-supervised learning~\cite{Park2020ImprovedNS}. The plot shows that if all utterances are selected, then a conventional system and an attention-based sequence-to-sequence model have similar performance (7$\sim$8\% WER). As the confidence threshold increases, the WER of the conventional system monotonically decreases. However, for the end-to-end model, a higher threshold does not always mean a reduced WER. The spike in \figdot{teaser} indicates that the end-to-end model is overconfident based on softmax probabilities.

To address the overconfidence issue, the \emph{confidence estimation module} (CEM) is proposed for attention-based sequence-to-sequence models in \sectdot{cem}. \twosect{setup}{exp} describes the setup and experiments that demonstrate the effectiveness of CEM. \sect{analysis} shows the generalisation performance and impact on downstream tasks of the proposed method. Conclusions are drawn in \sectdot{conclusion}.
\section{Confidence Estimation Module}
\begin{figure}[ht]
    \centering
    \includegraphics[width=0.95\linewidth]{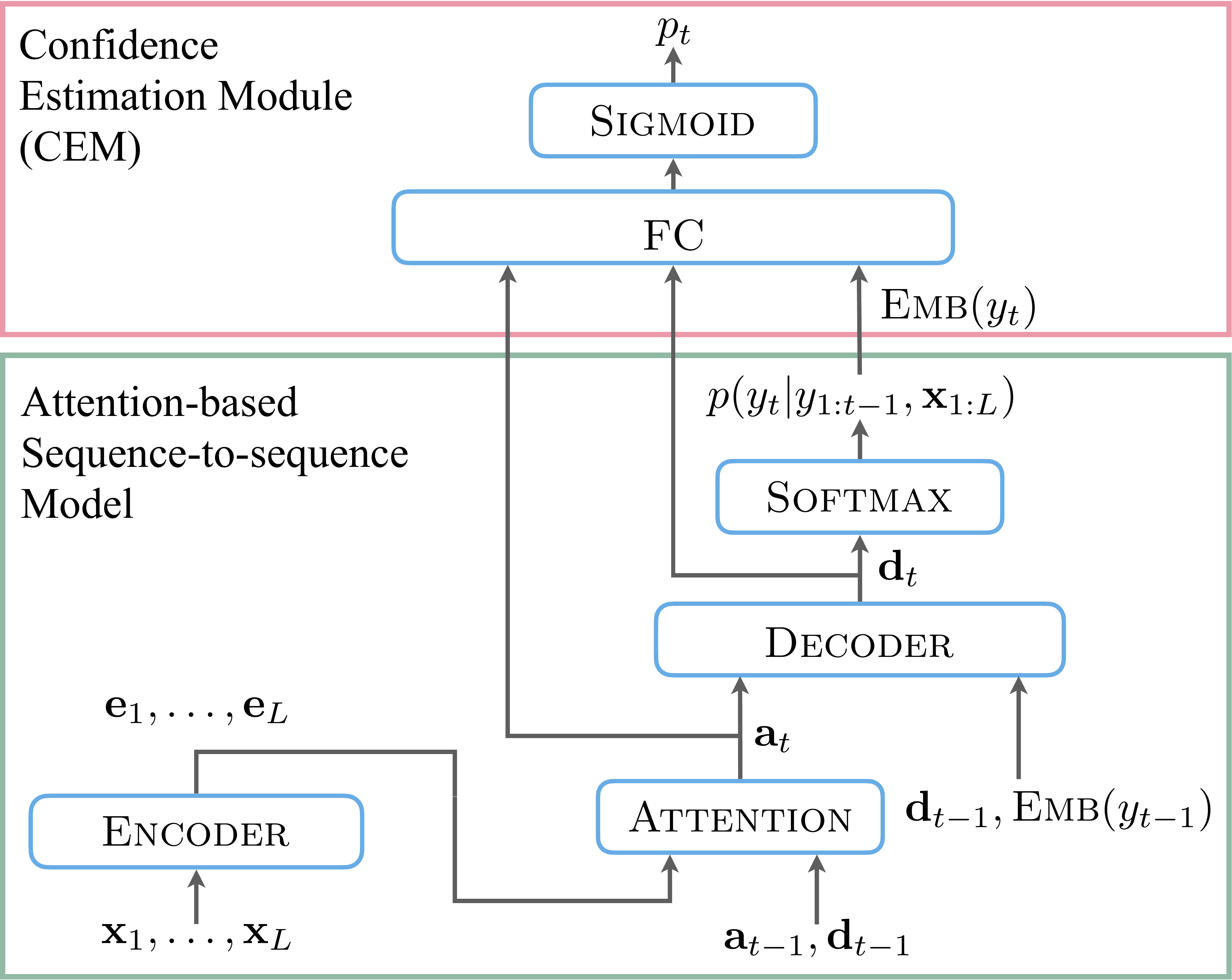}
    \vspace{-0.3em}
    \caption{Confidence estimation module (CEM) for attention-based sequence-to-sequence models.}
    \label{fig:cem}
    \vspace{-1em}
\end{figure}

\postsec
\label{sec:cem}
An attention-based sequence-to-sequence model, such as the Listen, Attend and Spell (LAS) model~\cite{Chan2016ListenAA}, generally consists of an encoder, an attention mechanism and a decoder. As shown in the green block in \figdot{cem}, for an input utterance $\mathbf{x}_1, \dots, \mathbf{x}_L$, the encoder first transforms the input feature for each frame
\begin{align}
    \mathbf{e}_{1:L} = \textsc{Encoder}(\mathbf{x}_{1:L}).
\vspace{-0.3em}
\end{align}
Then at a decoding step $t$,
\begin{align}
    \mathbf{a}_t & = \textsc{Attention}(\mathbf{a}_{t-1}, \mathbf{d}_{t-1}, \mathbf{e}_{1:L}), \\
    \mathbf{d}_t & = \textsc{Decoder}(\mathbf{a}_{t}, \mathbf{d}_{t-1}, \textsc{Emb}(y_{t-1})),\\
    p(y_t|y_{1:t-1},\mathbf{x}_{1:L}) & = \textsc{Softmax}({\mathbf{d}_t}).
\vspace{-0.3em}
\end{align}
where $y$ is a label token normally representing a grapheme or a word piece~\cite{Chiu2018StateoftheArtSR} for ASR tasks. At the token level, confidence scores for ASR are defined as the probability of the token being correct. If the recogniser is very confident about the output token, then the corresponding confidence score should be close to 1. Word or utterance level confidence scores can be obtained by taking the average of tokens within a word or an utterance. For attention-based sequence-to-sequence models, each step in the auto-regressive decoder is treated as a classification task over all possible output tokens. However, there is a subtle difference between calibration for standard classification and confidence scores for sequences. For a hypothesis sequence, each token can either be correct, a substitution or an insertion. Because of the auto-regressive nature of the decoder and the use of teacher forcing approach for training, calibration behaviour for sequences with an incorrect history is uncertain. Furthermore, the model can be poorly calibrated when the model becomes very deep and large in pursuit of state-of-the-art performance~\cite{Guo2017OnCO}.

To obtain high-quality confidence scores while keeping the model performance, the \emph{confidence estimation module} (CEM) is proposed as shown in the red block in \figdot{cem}. The CEM is designed to be a lightweight module that can be easily configured on top of any attention-based sequence-to-sequence model. The CEM gathers information from the attention mechanism, the decoder state and the current token embedding which is fed into a fully connected (FC) layer. Then the sigmoid output layer generates a value between 0 and 1 that indicates the confidence score $p_t$ for the current token, \ie
\begin{equation}
    p_t = \textsc{Sigmoid}(\textsc{FC}(\mathbf{a}_t, \mathbf{d}_t, \textsc{Emb}(y_t))).
    \vspace{-0.3em}
\end{equation}

Assuming there is an existing well-trained LAS model, n-best hypotheses can be generated by running a forward pass through the encoder and running a beam search through the decoder. Then the edit distance between each hypothesis sequence can be computed with respect to the ground truth reference sequence. The alignment from the edit distance computation can be used as the target for confidence if correct tokens are assigned as 1 while substituted or inserted tokens are assigned as 0. For example, if the ground truth sequence is ``\texttt{A B C D.}" and one of the hypotheses is ``\texttt{A C C D.}'', then the binary target sequence is $\mathbf{c}=[1, 0, 1, 1]$. For each of the n-best hypotheses, the CEM is trained to minimise the binary cross entropy between the estimated confidence $\mathbf{p}$ and the target $\mathbf{c}$. 
\vspace{-0.3em}
\begin{equation}
    \mathcal{L}(\mathbf{c},\mathbf{p}) = -\dfrac{1}{T}\sum_{t=1}^T \Big(c_t\log(p_t) + (1-c_t)\log(1-p_t)\Big).
    \vspace{-0.3em}
\end{equation}
The total loss for an utterance is the aggregated confidence estimation loss for all the n-best hypotheses. During training, all parameters for the attention-based sequence-to-sequence model are fixed.

\presec
\section{Experimental Setup}
\label{sec:setup}
\subsection{Evaluation Metrics}
\postsec
For each hypothesis, an alignment between the hypothesis and the corresponding ground truth can be obtained by computing the Levenshtein distance. If the hypothesis token is the same as the aligned reference token, the target confidence should be 1 and 0 otherwise.

To measure the quality of confidence scores, \emph{normalised cross-entropy} (NCE) is a frequently used metric~\cite{Siu1997ImprovedEE}. If we gather confidence scores for all tokens $\mathbf{p}=[p_1, \dots, p_N]$ where $p_n\in [0,1]$ and their corresponding target confidence $\mathbf{c}=[c_1, \dots, c_N]$ where $c_n\in\{0,1\}$, NCE is given by
\begin{equation}
    \text{NCE}(\mathbf{c},\mathbf{p}) = \dfrac{H(\mathbf{c}) - H(\mathbf{c},\mathbf{p})}{H(\mathbf{c)}}
    \vspace{-0.3em}
\end{equation}
where $H(\mathbf{c})$ is the entropy of the target sequence and $H(\mathbf{c},\mathbf{p})$ is the binary cross-entropy between the target and the estimated confidence scores. When confidence estimation is systematically better than the word correct ratio ($\sum_{n=1}^Nc_n/N$), NCE is positive. For perfect confidence scores, NCE is 1. NCE measures how close the confidence score is to the the probability of the recognised word being correct.

However, in applications such as keyword spotting, it is the order of tokens ranked by confidence scores that matters. In these cases, operating points have to be chosen where hypotheses with confidence scores above a certain threshold $\tilde{p}$ are deemed to be correct and incorrect otherwise. Precision-recall (P-R) curves are commonly used to illustrate the operating characteristics~\cite{Davis2006TheRB}.
\begin{equation}
    \text{precision}(\tilde{p}) = \dfrac{\text{TP}(\tilde{p})}{\text{TP}(\tilde{p})+\text{FP}(\tilde{p})},\;\; \text{recall}(\tilde{p}) =  \dfrac{\text{TP}(\tilde{p})}{\text{TP}(\tilde{p})+\text{FN}(\tilde{p})}.
\end{equation}
where TP is true positives, FP is false positives and FN is false negatives. Normally, when the threshold $\tilde{p}$ increases, there are fewer false positives and more false negatives, which leads to higher precision and lower recall. The trade-off behaviour between precision and recall yields a downward trending curve from the top left corner to the bottom right corner.
Therefore, the \emph{area under the curve} (AUC) can measure the quality of the confidence estimator, which has a maximum value of 1. It is worth noting that two confidence estimators can have the same AUC value but different NCE values. 

\presec
\subsection{Data}
\postsec
The LibriSpeech train-clean-100 subset~\cite{Panayotov2015LibrispeechAA} is used for model training. The 100-hour subset reflects a typical use case of confidence scores where the WERs are moderately high and the amount of supervised data is limited. The dev and test sets are dev-clean/dev-other and test-clean/test-other sets. The input features are 80-dimension filterbank coefficients with $\Delta$ and $\Delta\Delta$. The output targets are 16k word-pieces tokenised from the full LibriSpeech training set using a WPM model~\cite{Schuster2012JapaneseAK}. All experimental results in \sectdot{exp} on LibriSpeech are on the test-clean/test-other sets.

\presec
\subsection{Baseline Model}
\postsec
\label{ssec:baselinesetup}
The baseline model is an LAS model trained using the open-source Lingvo toolkit~\cite{Shen2019LingvoAM}. The encoder consists of a 2-layer convolutional neural network with max-pooling and stride of 2 and a 4-layer bi-directional long short-term memory (LSTM) network with 1024 units in each direction. The decoder has a 2-layer uni-directional LSTM network with 1024 units. The total number of parameters in the baseline model is 184 million. The optimiser is Adam with learning rate 0.001 and the batch size is 512. During training, five regularisation techniques have been adopted, including dropout~\cite{srivastava2014dropout} of 0.1 on the decoder, uniform label smoothing~\cite{Szegedy2016RethinkingTI} of 0.2, Gaussian weight noise~\cite{Graves2011PracticalVI} with zero mean and a standard deviation of 0.05 for all model parameters after 20k updates, SpecAugment~\cite{Park2019SpecAugmentAS} with 2 frequency masks with mask parameter $F=27$, 2 time masks with mask parameter $T = 40$, and time warping with warp parameter $W = 40$, and an exponential moving average (EMA)~\cite{Polyak1992AccelerationOS} of all model parameters were used during training. 
\presec
\section{Experimental Results}
\postsec
\label{sec:exp}
\subsection{Effect of Regularisation Methods on Confidence Scores}
\postsec
Many training techniques have been developed for deep neural networks. They normally reduce the extent to which the model overfits to the training data and improve generalisation of the model. The state-of-the-art performance is often achieved when a large model is trained using aggressive regularisation~\cite{Chiu2018StateoftheArtSR}. As for the baseline setup described in \ssectdot{baselinesetup}, five techniques are used. Broadly speaking, regularisation methods can be classified into three categories, \ie augmenting input features (SpecAugment), manipulating model weights (dropout, EMA \& weight noise), and modifying output targets (label smoothing).
\vspace{-0.5em}
\begin{table}[ht]
    \centering
    \begin{tabular}{l|rrr}
        \toprule
         & WER $\downarrow$ & AUC $\uparrow$ & NCE $\uparrow$ \\
        \midrule
        baseline & 7.5/21.6 & 0.976/0.912 & -0.195/ 0.131\\
        \midrule
        \;$-$ dropout & 7.8/22.0 & 0.977/0.916 & -0.204/ 0.130\\
        \;$-$ EMA & 8.2/24.8 & 0.974/0.903 & -0.189/ 0.120\\
        \;$-$ label smoothing & 10.6/24.6 & 0.985/0.950 & 0.106/-0.131\\
        \;$-$ weight noise & 12.9/25.8 & 0.978/0.925 & -0.459/-0.012 \\
        \;$-$ SpecAugment & 10.8/34.3 & 0.952/0.911 & 0.012/ 0.160\\
        \bottomrule
    \end{tabular}
    \caption{ASR and token-level confidence performance by removing a regularisation method from the baseline model on test-clean/test-other. Confidence scores are based on softmax probabilities.}
    \vspace{-1em}
    \label{tab:regularisation}
\end{table}

In \tbl{regularisation}, the ASR results and confidence performance of five additional models are shown, where each model is trained by removing one regularisation technique from the baseline setup. As expected, removing any of these regularisation methods results in an increased WER. However, the confidence performance may not worsen when a regularisation method is excluded. Also, AUC and NCE values do not always change in the same direction. Although disentangling the contribution of each technique is challenging, \tbl{regularisation} shows that augmenting input features and manipulating model weights can generally improve the confidence performance for at least one metric. Interestingly, by removing label smoothing, AUC is significantly better than the baseline and the gap of NCE values between test-clean and test-other sets is reversed. This shows that although softmax probabilities can be directly used as confidence scores, they can be heavily affected by regularisation techniques, especially if output targets are modified. 

\presec
\subsection{Confidence Estimation Module}
\postsec
\label{ssec:cem}
To keep good ASR performance while having reliable confidence scores, a dedicated CEM can be trained as in \sectdot{cem}. The CEM is added on top of the baseline model. During the CEM training, more aggressive SpecAugment (10 time masks with mask parameter $T = 50$) is used to increase the WER on the training set for more negative training samples. For each utterance, 8-best hypotheses are generated on-the-fly and are aligned with the reference to obtain the binary training targets. The CEM only has one fully-connected layer with 256 units. The number of additional parameters is 0.4\% of the baseline LAS model.
Piece-wise linear mappings (PWLMs)~\cite{Evermann2000LargeVD} are estimated on dev-clean/dev-other and are then applied to test-clean/test-other, so that the confidence scores better match with token or word accuracy. Since PWLM is monotonic, NCE is boosted while AUC remains unchanged as the relative order of confidence scores is unchanged.

\vspace{-0.5em}
\begin{table}[ht]
    \centering
    \begin{tabular}{cc|rrr}
        \toprule
         & & \multirow{2}{*}{AUC $\uparrow$} & NCE $\uparrow$ & NCE $\uparrow$ \\
         & & & (w/o PWLM) & (w/ PWLM)\\
        \midrule
        \multirow{2}{*}{token} & softmax & 0.976/0.912 & -0.195/0.131 & 0.166/0.172\\
         & CEM & \bf{0.990/0.958} & 0.189/0.019 & \bf{0.344/0.275}\\
        \midrule
        \multirow{2}{*}{word} & softmax & 0.981/0.927 & -0.180/0.139 & 0.269/0.195\\
         & CEM & \bf{0.990/0.962} & 0.192/0.039 & \bf{0.350/0.270}\\
        \bottomrule
    \end{tabular}
    \caption{Comparison of confidence scores between using softmax probabilities and using the CEM on the baseline model. The first row corresponds to the baseline in \tbl{regularisation}.}
    \label{tab:cem}
    \vspace{-0.5em}
\end{table}

\tbl{cem} reports the confidence metrics at the token level and the word level. The word-level confidence is the average of the token-level ones if a word consists of multiple tokens.
The AUC is improved at both token and word levels by using CEM. Unlike softmax probabilities, NCE values are all positive for CEM. After PWLM, CEM yields much higher NCE values than softmax probabilities.
\vspace{-0.5em}
\begin{figure}[ht]
     \centering
     \begin{subfigure}{0.45\linewidth}
         \centering
         \includegraphics[width=\linewidth]{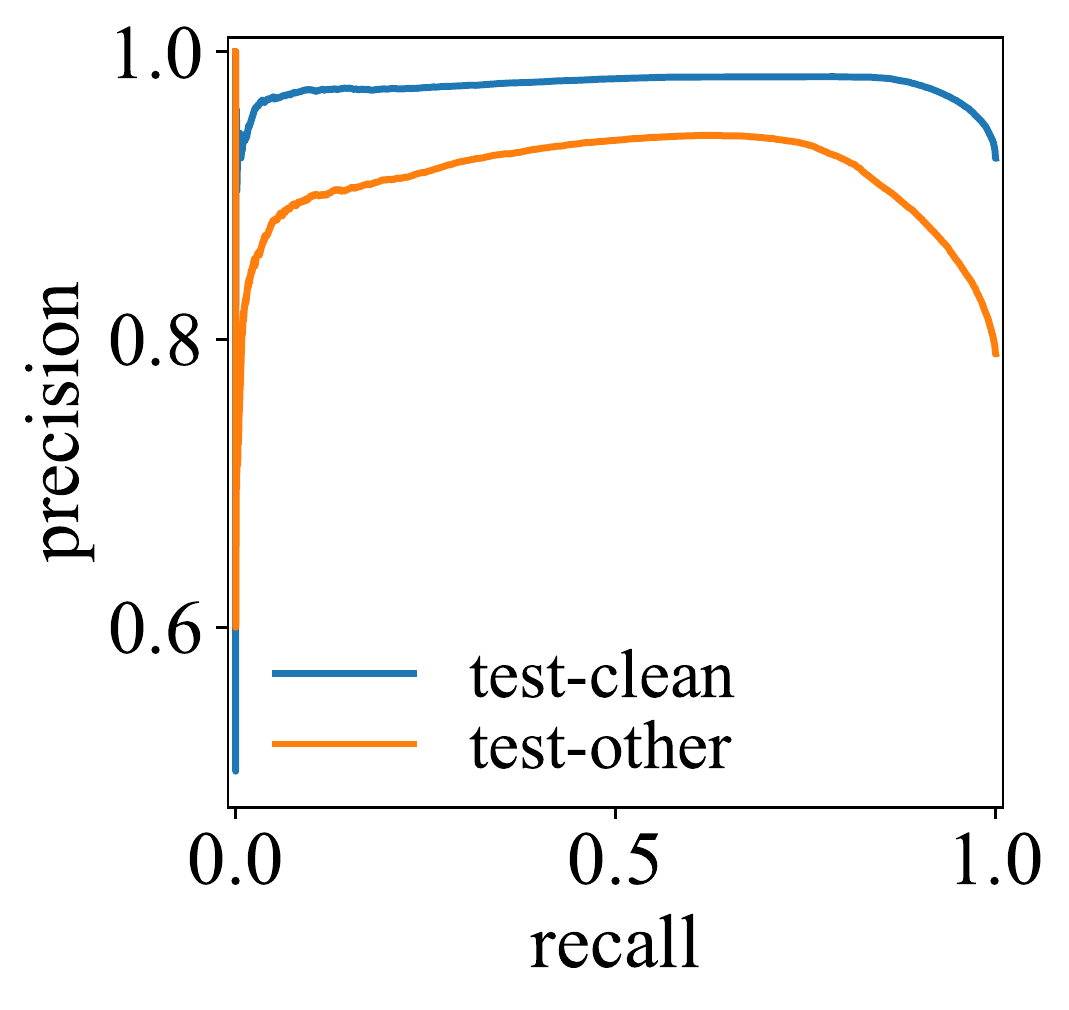}
         \vspace{-1.5em}
         \caption{softmax}
         \label{fig:softmax_pr}
     \end{subfigure}
     \hfill
     \begin{subfigure}{0.45\linewidth}
         \centering
         \includegraphics[width=\linewidth]{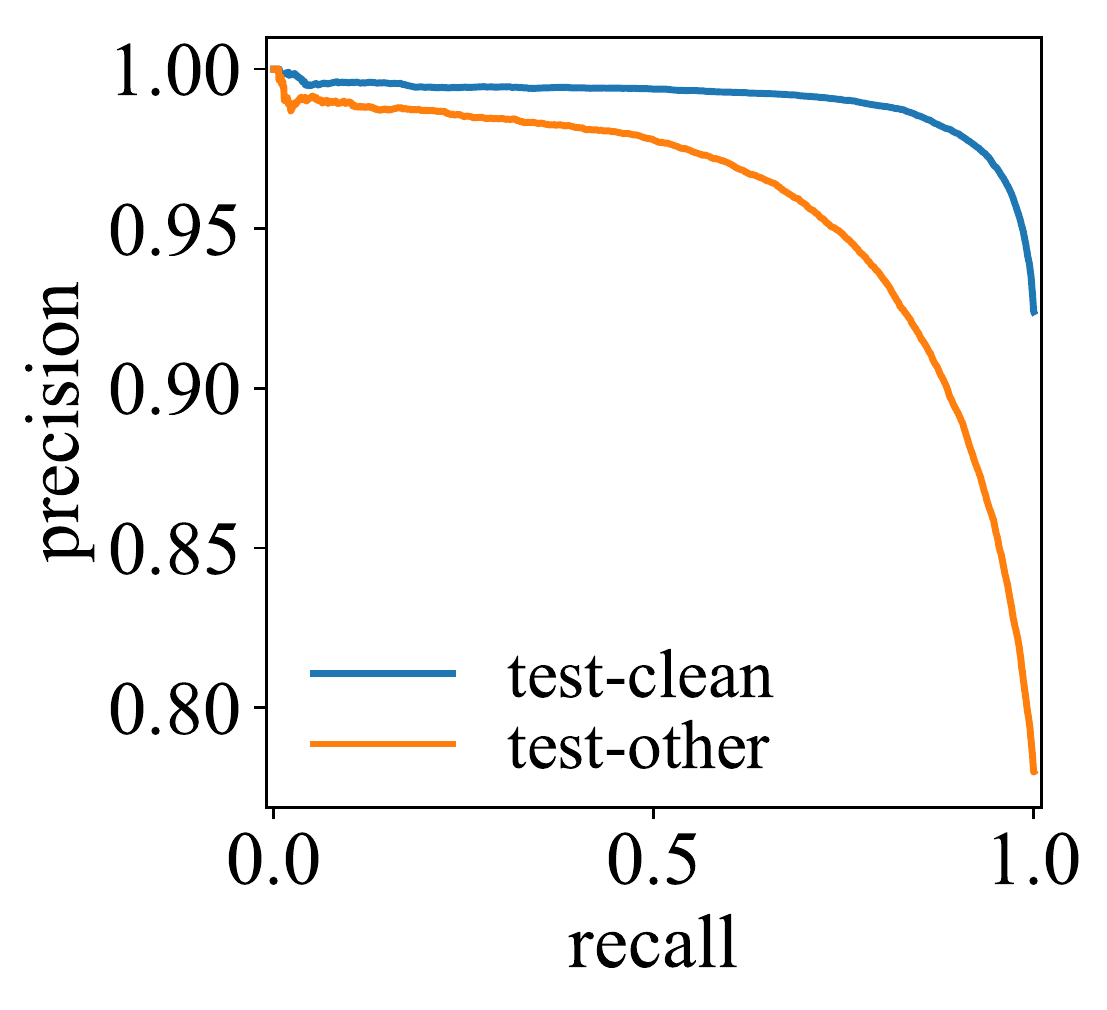}
         \vspace{-1.5em}
         \caption{CEM}
         \label{fig:cem_pr}
     \end{subfigure}
     \vspace{-1em}
    \caption{Precision-recall curves for token-level confidence scores on LibriSpeech test-clean and test-other sets.}
    \vspace{-0.5em}
    \label{fig:pr}
\end{figure}

However, AUC values do not show the whole picture. As shown in \figdot{pr}, the P-R curves of softmax and CEM are drastically different. A sharp downward spike at the high-confidence region in \figdot{softmax_pr} corresponds to a low precision and a low recall. In other words, softmax probabilities are overconfident for some incorrect tokens, which also explains the spike shown in \figdot{teaser}. CEM, however, suffers little from overconfidence as \figdot{cem_pr} depicts the desired trade-off between precision and recall. Overall, the CEM is a more reliable confidence estimator under both metrics.

\presec
\subsection{Effect of Language Model Fusion on Confidence Scores}
\postsec
\label{ssec:fusion}
For end-to-end ASR, shallow fusion of a language model (LM)~\cite{Glehre2015OnUM} is commonly used to improve ASR performance during decoding.
The effect of shallow fusion on confidence estimation is investigated. The language model used is a three-layer LSTM network with width 4096 trained on the LibriSpeech LM corpus, which shares the same word-piece vocabulary as the attention-based sequence-to-sequence model.
To take LM information for confidence estimation, the input to the CEM is extended by the LM probability for the current token. Other setups are the same as \ssectdot{cem}.
\vspace{-0.5em}
\begin{table}[ht]
    \centering
    \begin{tabular}{lc|rrr}
        \toprule
         & & WER $\downarrow$ & AUC $\uparrow$ & NCE $\uparrow$ \\
        \midrule
        \multirow{2}{*}{baseline} & softmax & \multirow{2}{*}{7.5/21.6} & 0.981/0.927 & 0.269/0.195\\
        & CEM & & \bf{0.990/0.962} & \bf{0.350/0.270}\\
        \midrule
        \multirow{2}{*}{\;$+$ LM} & softmax & \multirow{2}{*}{6.8/19.8} & 0.981/0.928 & 0.103/0.109\\
        & CEM & & \bf{0.991/0.966} & \bf{0.337/0.263}\\
        \bottomrule
    \end{tabular}
    \caption{ASR and word-level confidence performance for models with and without RNNLM shallow fusion (with PWLM).}
    \vspace{-0.5em}
    \label{tab:lm}
\end{table}

\tbl{lm} shows the word-level confidence scores after PWLM for both softmax probabilities and CEM. Although WERs on test-clean and test-other decreased by 8$\sim$9\% relatively, there is no clear improvement on AUC and even substantial degradation for NCE. By comparing the first and second blocks of \tbl{lm}, the CEM improves the quality of confidence estimation more significantly when an additional LM is used. The contrast of P-R curves between softmax and CEM with LM shallow fusion is similar to \figdot{pr}.
\presec
\section{Analysis}
\postsec
\label{sec:analysis}
\subsection{Generalisation to a Mismatched Domain}
\postsec
Since CEM is a model-based approach and the training data for CEM is the same as the ASR, the model is naturally more confident on the training set. Although the mismatch between training and test for CEM is mitigated by having more aggressive augmentation during training and applying PWLMs estimated on dev sets during testing, it is still unclear how well the confidence scores from CEM generalises to data from a mismatched domain.
\vspace{-0.5em}
\begin{table}[ht]
    \centering
    \begin{tabular}{lc|ccc}
        \toprule
         & & WER $\downarrow$ & AUC $\uparrow$ & NCE $\uparrow$ \\
        \midrule
        \multirow{2}{*}{baseline} & softmax & \multirow{2}{*}{18.7} & 0.935 & 0.230\\
        & CEM & & \textbf{0.970} & \textbf{0.280}\\
        \midrule
        \multirow{2}{*}{\;$+$ LM} & softmax & \multirow{2}{*}{17.7} & 0.933 & 0.159\\
        & CEM & & \textbf{0.965} & \textbf{0.266}\\
        \bottomrule
    \end{tabular}
    \caption{ASR and confidence performance on WSJ eval92. PWLM is estimated on LibriSpeech dev-other set. LM used is trained on LibriSpeech LM corpus as in \ssectdot{fusion}.}
    \vspace{-1em}
    \label{tab:wsj}
\end{table}

Wall Street Journal (WSJ)~\cite{Paul1992TheDF}, a dataset of clean read speech of news articles, is in a moderately mismatched domain compared to LibriSpeech in terms of speaker, style and vocabulary. In \tbl{wsj}, WSJ eval92 test set is fed into the same setup as in \ssectdot{fusion}, where all models are trained on LibriSpeech. Similar to observations in \tbl{lm}, shallow fusion worsens the confidence estimation by softmax probabilities despite reduced WER. CEM improves the quality of confidence estimation significantly with or without LM.



\presec
\subsection{Implications for Downstream Tasks}
\postsec
Confidence scores are widely used to select unlabelled data for semi-supervised learning~\cite{Park2020ImprovedNS} to improve ASR performance. First, a speech recogniser is trained using the limited transcribed data. Then the recogniser transcribes the unlabelled data, which can be taken as noisy labels to train the existing model further. However, erroneous automatic transcription can hurt the model. 
Assuming confidence scores can reflect the word error rate well, filtering out utterances with low confidence can be beneficial to semi-supervised training.
\vspace{-0.5em}
\begin{figure}[ht]
     \centering
     \begin{subfigure}{0.45\linewidth}
         \centering
         \includegraphics[width=\linewidth]{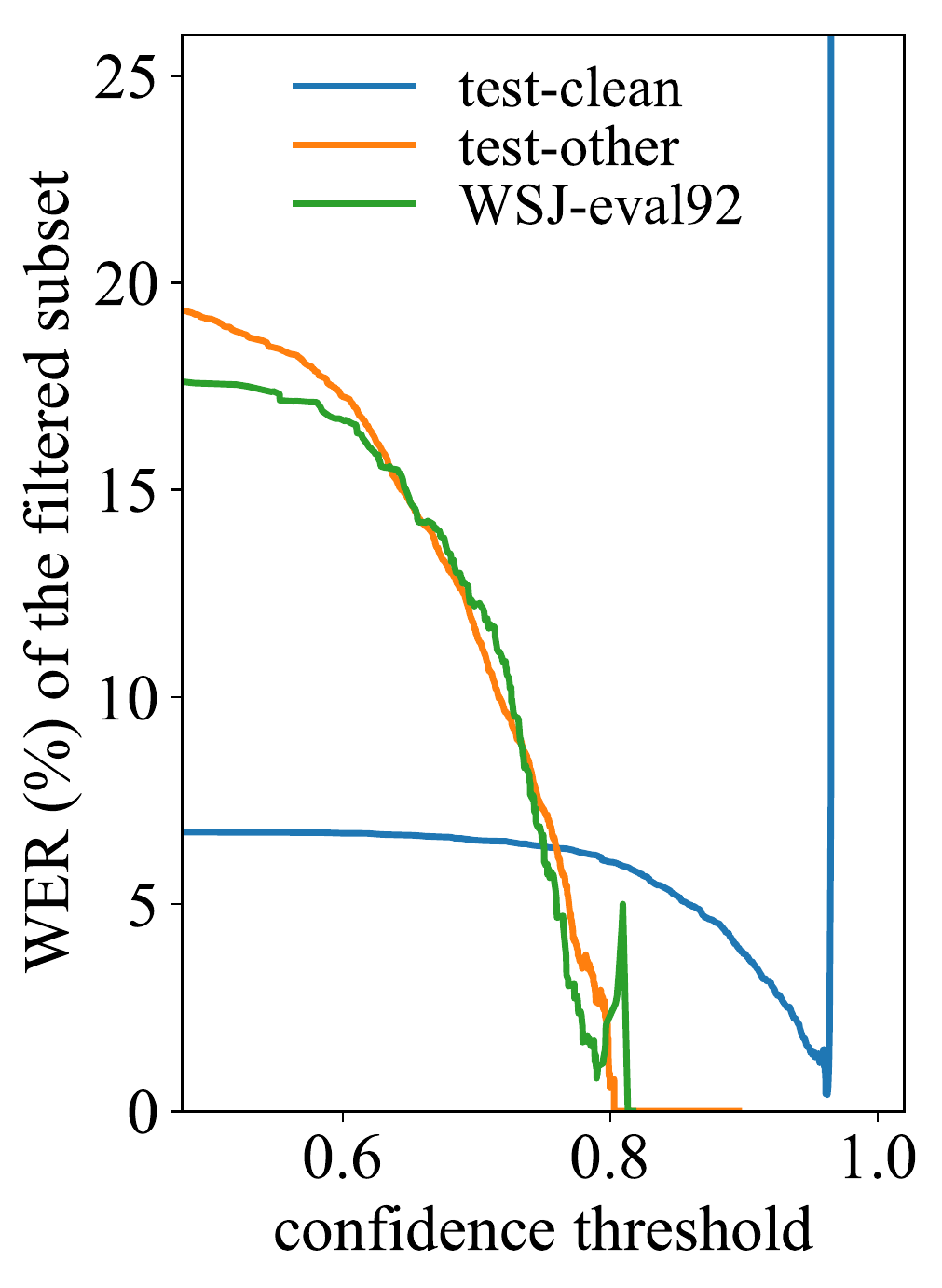}
         \vspace{-1.5em}
         \caption{softmax}
         \label{fig:softmax_filter}
     \end{subfigure}
     \hfill
     \begin{subfigure}{0.45\linewidth}
         \centering
         \includegraphics[width=\linewidth]{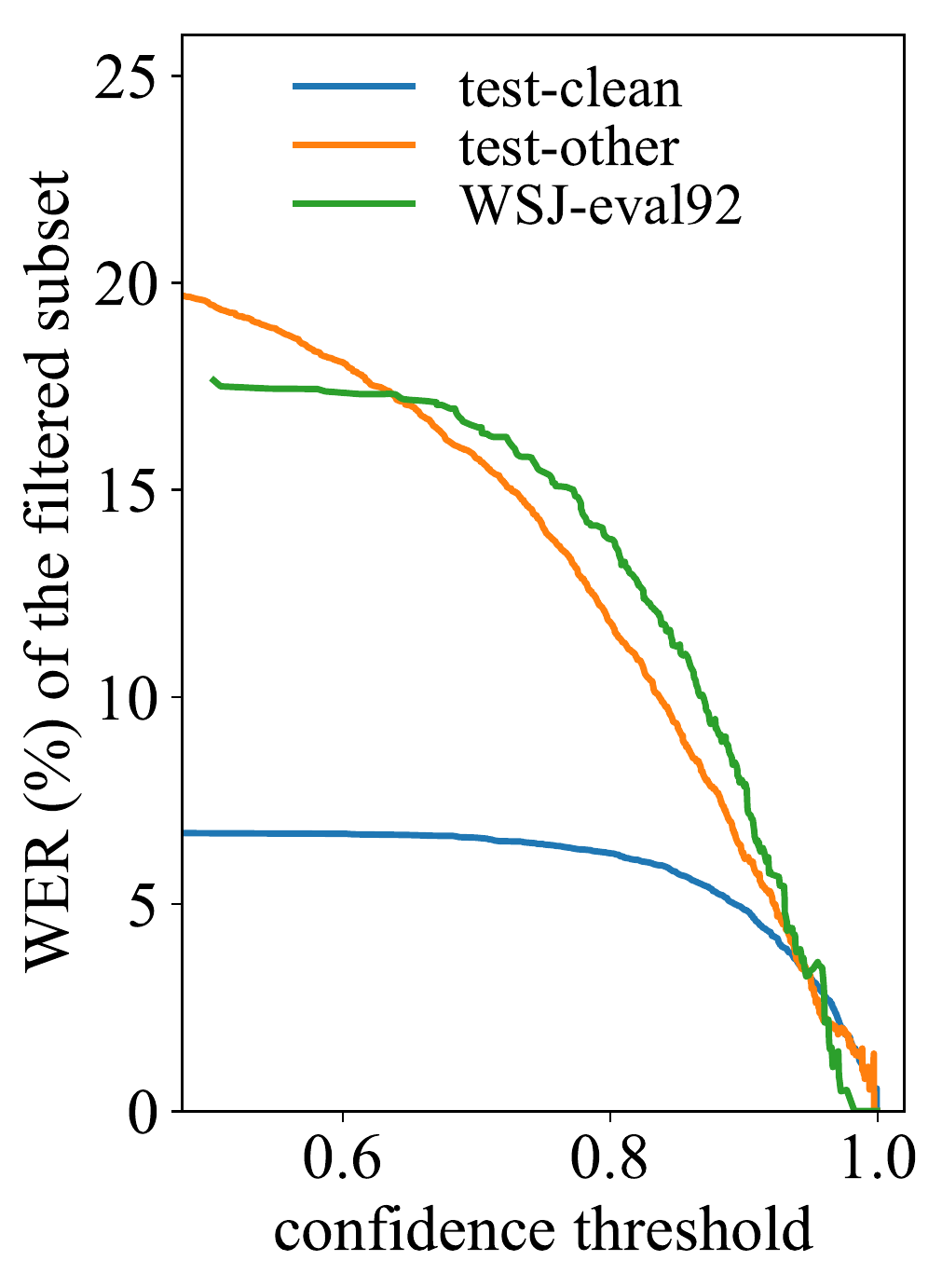}
         \vspace{-1.5em}
         \caption{CEM}
         \label{fig:cem_filter}
     \end{subfigure}
     \vspace{-0.5em}
    \caption{WERs of filtered utterances \wrt confidence thresholds for softmax and CEM with LM shallow fusion.}
    \label{fig:filter}
    \vspace{-1em}
\end{figure}

Similar to \figdot{teaser} and plots used for semi-supervised learning~\cite{Park2020ImprovedNS}, \figdot{filter} shows the WER of the filtered utterances whose confidence scores are above the corresponding threshold. If confidence scores strongly correlate with WER, a higher threshold will filter a subset with lower WER. In \figdot{softmax_filter}, sharp spikes at the high confidence region clearly indicates overconfidence based on softmax probabilities. In contrast, curves for all three test sets in \figdot{cem_filter} are monotonically going down without spikes, which shows that confidence scores by CEM match WER much more closely.
\presec
\section{Conclusions}
\postsec
\label{sec:conclusion}
Using softmax probabilities of attention-based sequence-to-sequence models for confidence estimation is not reliable. To this end, the lightweight confidence estimation module (CEM) is proposed. The effectiveness of CEM is demonstrated with and without shallow fusion of a language model. Furthermore, the performance of CEM can generalise well on a slightly mismatched domain and may benefit various downstream tasks, such as adaptation~\cite{Uebel2001SpeakerAU}, semi-supervised training~\cite{Park2020ImprovedNS} and model combination~\cite{Li2019IntegratingSA,Fiscus1997APS}.
A similar module to CEM can also be introduced for deletion prediction as deletion errors are not within the scope of confidence estimation but can be important for many applications~\cite{Seigel2014DetectingDI,Ragni2018ConfidenceEA}. With minimal modification, CEM is also applicable to recurrent neural network transducer (RNN-T) for speech recognition~\cite{Graves2012SequenceTW}.


\newpage
\section{References}
\vspace{-0.3em}
\begingroup
\renewcommand{\section}[2]{}
\bibliographystyle{IEEEbib}
\bibliography{refs}
\endgroup
\end{document}